# Sensing Small Changes in a Wave Chaotic Scattering System

Biniyam Tesfaye Taddese[1], Thomas M. Antonsen[1,2], Edward Ott[1,2], and Steven M. Anlage[1,2]

[1]Department of Electrical and Computer Engineering, University of Maryland, College Park, Maryland 20742-3285, USA
[2]Department of Physics, University of Maryland, College Park, Maryland 20742-4111, USA

Classical analogs of the quantum mechanical concepts of the Loschmidt Echo and quantum fidelity are developed with the goal of detecting small perturbations in a closed wave chaotic region. Sensing techniques that employ a one-recording-channel time-reversal-mirror, which in turn relies on time reversal invariance and spatial reciprocity of the classical wave equation, are introduced. In analogy with quantum fidelity, we employ Scattering Fidelity techniques which work by comparing response signals of the scattering region, by means of cross correlation and mutual information of signals. The performance of the sensing techniques is compared for various perturbations induced experimentally in an acoustic resonant cavity. The acoustic signals are parametrically processed to mitigate the effect of dissipation and to vary the spatial diversity of the sensing schemes. In addition to static boundary condition perturbations at specified locations, perturbations to the medium of wave propagation are shown to be detectable, opening up various real world sensing applications in which a false negative cannot be tolerated.

Detecting small changes inside enclosures with complicated boundary conditions can be of practical importance. The small changes inside such enclosures can either be perturbations of the boundary conditions or the medium of wave propagation. Examples of practical situations where such sensitive detection capabilities are beneficial include the following: strict surveillance of the interior of an unoccupied building, scrutiny of a potentially harmful re-arrangement of objects inside an enclosure being transported, supervision of a tightly sealed chamber for gas leaks, inspection of a confined fluid for hazardous turbulence ,etc. In each of these circumstances, false negatives may not be tolerated and it is essential to have a sensitive detection mechanism with broad spatial coverage.

A traditional approach of monitoring a complicated enclosure is to use a network of several wave-based sensor units each monitoring a limited region of the enclosure. Our approach is to use a single, cost effective wave based sensor unit that can monitor the complicated enclosure as a whole. Unlike traditional sensors, the sensor is not confounded by multiple reflections. Instead it actually takes advantage of, and works better using, the information of ray trajectories that ergodically explore the cavity through multiple reflections before collapsing back onto the sensor.

## I. INTRODUCTION

In the limit where the wavelength is small compared to the characteristic size of the enclosure, wave propagation inside the enclosure can be modeled using ray trajectories. The irregularities in the boundaries of the enclosure results in sensitive dependence of the trajectories of the rays on their initial conditions. This property is known as "ray chaos". As usually defined, chaos is a property associated with nonlinear dynamical systems, and linear wave systems cannot be chaotic [1]. However, wave systems whose classical (small wavelength) limit is ray chaotic show interesting properties. The study of such wave systems is called "wave chaos" or "quantum chaos" [2]. In related work, we have created a random coupling model to understand the frequency-domain and time-domain properties of wave chaotic systems [3-6], and this model has been tested through experiments on a microwave resonator [7-8].

The underlying ray chaos in a wave chaotic system promises to be useful in detecting small changes to the system. In this paper, two classes of sensing techniques, which take advantage of the sensitive dependence of wave trajectories on small changes to the system, are studied. The first class of sensing techniques is based on a "propagation comparison" of two distinct wave excitations of the system. The second class of sensing techniques exploits time reversal invariance and spatial reciprocity of the wave equation; it works by comparing pulses reconstructed using a time reversal of the wave excitations of the system. These sensing techniques are tested experimentally, and their performance under various circumstances is compared quantitatively.

In this paper, the quantum mechanical concepts of fidelity and Loschmidt Echo (LE) are extended to classical waves with the goal of sensing perturbations to a scattering environment. The physical theory behind these quantum mechanical concepts is briefly discussed in Section II. Section III is a summary of the literature in related areas. The operation of four different acoustic sensing techniques tested in an enclosed stairwell is explained in Section IV. In this section, an indicator value of perturbation is defined for each sensing technique. The details of signal processing done to mitigate the effect of dissipation, and to alter the spatial range sensitivity of the sensors is also included in Section IV. Section V explains a method to standardize the indicator values of perturbation of the different sensing techniques to enable consistent comparisons. The performance of the sensing techniques for perturbations made at different locations in the stairwell is summarized in Section VI. Section VII contains some comments on the relative merits of these different sensing techniques and discusses some of the experimental limitations. Finally, a brief conclusion is presented in Section VIII. This paper expands considerably on a preliminary publication [9].

## II. THEORY

Wave chaotic systems have wave scattering properties that are quite sensitive to small perturbations of the scattering environment. One can define two mathematically equivalent measures of this sensitivity in the context of quantum mechanics; these are the quantum fidelity and the LE [10, 11]. Each of these mathematically equivalent quantities measures the sensitivity of the dynamics of a quantum mechanical system to small perturbations of its Hamiltonian.

The LE can be defined as follows. A system is prepared in a given initial state $|\Psi(0)>$, propagated forward in time under an unperturbed time-reversible Hamiltonian $H$ to some time $t$, $|\Psi(t)>=U(t)|\Psi(0)>$ where $U(t)=\exp(-iHt/\hbar)$ is the time evolution operator. At that time the evolution is stopped and the Hamiltonian is perturbed by a small amount $H'$, so that $H \rightarrow H+H'$. The system is then propagated backward in time under the perturbed Hamiltonian $H+H'$ to create another state $U'(-t)U(t)|\Psi(0)>$ where $U'(-t)=\exp[i(H+H')t/\hbar]$. The overlap of this forward and backward propagated state with the initial state is known as the LE, $LE_{H'}(t)=<\Psi(0)|U'(-t)U(t)|\Psi(0)>$.

The formula above for the LE can also be interpreted as the overlap of two different final states of the system which started out from the same initial state, $\Psi(0)$, but have been propagated forward in time with different Hamiltonians, namely $H$ and $H+H'$. Such a different interpretation of the same quantity defines the quantum fidelity. The quantum fidelity is unity in the absence of perturbations (i.e. $H'=0$) for any $H$ and $t$. However, in the presence of perturbations the quantum fidelity will decay with $t$ at a rate depending on $H$ and the perturbation. It is worth noting that despite their mathematical equivalence the implementation details of the computation or measurement of these quantities can be quite different, as we shall see below.

The theoretical equivalence of the LE and quantum fidelity motivates the exploration of their classical wave analogs with the goal of developing a practical perturbation sensor. In this paper we experimentally investigate two classes of sensing techniques which extend these two quantum mechanical concepts to classical waves. The paper devises a tunable sensor that overcomes the effects of dissipation in classical waves, and as a consequence, also creates a sensor with adjustable spatial range coverage. A statistical Figure of Merit is defined to compare the relative merits of the different sensing techniques developed. The Figure of Merit defined also helps to choose an optimum set of parameters for sensing a given perturbation.

The classical wave analog of the LE is implemented using a time reversal procedure which involves the following steps. Suppose that there is a cavity whose response to incident input signals can be characterized by a linear, causal, time invariant system. Let the reflected system response to an incident impulse be *s(t)*; the corresponding Fourier Transform of the impulse response (i.e. the transfer function) is denoted by $\hat{s}(\omega)$, which is a function of the Fourier frequency transform variable $\omega$ (in what follows we consider $\omega$ to be real). The first step of the time reversal procedure is to inject a narrow band, pulse modulated, incident input signal *a(t)* into the system and to retrieve the resulting reflected output *b(t)*. The Fourier Transforms of these signals obey the relation $\hat{b}(\omega) = \hat{s}(\omega)\hat{a}(\omega)$, where, because *b(t)* and *a(t)* are real, $\hat{s}^*(\omega) = \hat{s}(-\omega)$. After recording *b(t)*, consider time reversing it and reinjecting it as an incident signal *b(-t)*; the Fourier Transform of *b(-t)* is $\hat{b}(-\omega)$. The system's response to *b(-t)* is denoted by *b'(t)*. The Fourier Transform of *b'(t)* is given by $\hat{b}'(\omega) = \hat{s}(\omega)\hat{b}(-\omega) = \hat{s}(\omega)\hat{s}(-\omega)\hat{a}(-\omega) = |\hat{s}(\omega)|^2 \hat{a}(-\omega)$. This expression is examined for different loss mechanisms in the system as follows.

For a lossless system, the scattering transfer function obeys the relation $|\hat{s}(\omega)|^2 = 1$. Thus, in the lossless case, $\hat{b}'(\omega) = \hat{a}(-\omega)$ holds, which implies that *b'(t)=a(-t)*. This means that a time reversed version of the original input, *a(t)*, is recovered after *b(-t)* is injected into a lossless system. Thus, for the lossless case, the classical analog of the LE is unity, and the time reversal procedure described here is 'perfect'.

For a system that is lossy, $|\hat{s}(\omega)|^2$ generally depends on ω. As a result, the exact time reversed version of the OP is not expected to be reconstructed for the lossy case. This result will be used to justify the experimental imperfection of the time reversal procedure explained in Section IV.2.2.

Next, consider a special case of a lossy system which has uniform loss. To motivate the definition of uniform loss, first consider a lossless situation in which temporally sinusoidal waves inside the scattering region are described by the wave equation $[\nabla^2 + (\omega/v)^2]\Psi = 0$, where *v* is the wave velocity, and the dependent variable $\Psi$ is subject to a lossless $\omega$-independent boundary condition on the boundaries of the scattering region. In this lossless case, the assumed solution to the scattering problem is described by a scattering coefficient $\hat{s}_0(\omega)$, where $|\hat{s}_0(\omega)|^2 = 1$; here, the subscript zero denotes the lossless case. Now assume that loss is added uniformly in space to the medium, but not to the boundary conditions. For small loss and a wide range of loss mechanisms, this modifies the wave equation within the scattering region via the replacement $\omega \to \omega + i\gamma$. Furthermore, we assume that any $\omega$-dependence of the loss rate $\gamma$ is negligible within the frequency bandwidth of the incident pulse *a(t)*. Since the only $\omega$-dependence of the scattering problem is assumed to occur in the wave equation, the transfer function of the uniformly lossy system, $\hat{s}(\omega)$, is given by $\hat{s}(\omega) = \hat{s}_0(\omega + i\gamma)$. Therefore, for the uniform loss case, the Fourier Transform of *b'(t)* defined above is given by $\hat{b}'(\omega) = \hat{s}_0(\omega + i\gamma)\hat{s}_0[-(\omega + i\gamma)]\hat{a}(-\omega)$. Once again, $\hat{s}_0(\omega + i\gamma)\hat{s}_0[-(\omega + i\gamma)]$ generally depends on ω, and hence the time reversal procedure is not expected to work perfectly even in the uniform loss case; in other words that *b'(t)* is generally different from *a(-t)*.

However, we now argue that, if the uniform loss in the system is compensated by applying a proper time exponential amplification to *b(t)*, the time reversal procedure will still work. The exponential amplification involves multiplying *b(t)* by $e^{2\gamma t}$. The time reversed version of this exponentially amplified signal is $b(-t)e^{-2\gamma t}$, with a corresponding Fourier Transform $\hat{b}[-(\omega + i2\gamma)]$. The Fourier Transform of the response of the system to $b(-t)e^{-2\gamma t}$ is given by $\hat{b}'(\omega) = \hat{s}(\omega)\hat{b}[-(\omega + i2\gamma)]$. Here, $\hat{s}(\omega)$ can be written as $\hat{s}_0(\omega + i\gamma)$ and $\hat{b}[-(\omega + i2\gamma)]$ can be written as $\hat{s}[-(\omega + i2\gamma)]\hat{a}[-(\omega + i2\gamma)]$. After substituting these expressions and simplifying we get $\hat{b}'(\omega) = \hat{s}_0(\omega + i\gamma)\hat{s}_0[-(\omega + i\gamma)]\hat{a}[-(\omega + i2\gamma)]$. The expression $\hat{s}_0(\omega + i\gamma)\hat{s}_0[-(\omega + i\gamma)]$ is identically one for $\gamma$=0 and all $\omega$ as this is the lossless case. For arbitrary $\gamma$ we note that the product is an analytic function of $\omega$. Thus, by analytic continuation it is also equal to one for any $\gamma$. Therefore, $\hat{b}'(\omega) = \hat{a}[-(\omega + i2\gamma)]$. In the time domain, $b'(t) = a(-t)e^{-2\gamma t}$. If the time duration of the original input signal, *a(t)*, is short compared with $1/\gamma$, then , $b'(t) \approx a(-t)$ . Therefore, if the loss in the system is uniform, then the time reversal procedure is expected to approximately work with the help of the exponential amplification. This is our motivation to use exponential amplification, described in Section IV.3, assuming that the loss in the system roughly approximates the case of a uniform loss over the bandwidth of the original input signal.

While uniform loss does not strictly apply when there are reflection losses at the boundaries (generally these depend on angle of incidence), we still might expect that the uniform loss case applies approximately. To justify this expectation, we think of $\hat{s}(\omega)$ as resulting from multiple ray paths originating from the port and then returning to it after following paths that bounce from the scatterer boundaries multiple times. Insofar as the loss over such a path is approximately proportional to the path length (travel time), the uniform loss approximation is expected to apply. Furthermore, if these paths are long and involve many reflections, their complicated, chaotic, nature implies that the net reflection loss would involve an average of the losses over many different incidence angles of the rays on the boundary. Thus, approximately ergodic behavior of chaotic rays implies a self-averaging process over different incidence angles and approximately uniform loss for long ray paths.

## III. PREVIOUS RELATED WORK

The idea of quantifying perturbations to a system using either a "propagation comparison" of two different final states of the system obtained from a given initial state, or a comparison of an initial state with a final state of the system obtained by a time reversal mirror has been explored previously. The concept of quantum fidelity which quantifies the sensitivity of the dynamics of a quantum mechanical system to small perturbations of its Hamiltonian is well developed [10, 11]. The LE makes connection to spin-echo experiments widely used in nuclear magnetic resonance [12].

The concept of the LE has been extended to classical waves using "time-reversal mirrors" for acoustics [13, 14] and electromagnetics [15-17]. Ideally, time-reversal mirrors operate by collecting and recording a propagating wave as a function of time, and at some later time they propagate it in the opposite direction in a time-reversed manner. Experimentally, it is not generally possible to mirror all waves in this manner. Experimental time-reversal mirrors can however be realized in the special case of confined systems with highly reflective walls (so called 'billiard' systems) and classically chaotic ray dynamics such as those considered here. Under these conditions a single-channel time-reversal mirror can very effectively approximate the conditions required to measure the LE using classical waves [17,18]. The experimental set up for the measurement of the LE can be further simplified by exploiting the spatial reciprocity of the wave equation [9, 19]. Time-reversal mirrors have found a wide range of practical applications such as crack imaging in solids [20], and improved acoustic communication in air [21], among other things. Recently, it was proposed that time reversal mirrors could also be applied to quantum systems. [22]

On the other hand, the concept of quantum fidelity has been applied to classical waves as in the study of the Scattering Fidelity (SF) of acoustic waves, which is, practically speaking, the correlation between signals as a function of time [23-27]. The relative merits of the cross correlation and mutual information of acoustic signals in the context of underwater source detection has been studied, for example, in Ref. [28].

## IV. EXPERIMENTS

The goal of our experiment is to test the sensitivity of different sensing techniques to small perturbations of a monitored acoustic cavity. A two story tall stairwell of dimensions 6m deep x 2.5m wide x 6.5m high serves as our enclosure under surveillance. [See Figure 1] A Samson C01U microphone and a desktop computer speaker that are about 1m apart are set up inside the stairwell, and are controlled by a laptop computer that is stationed outside the enclosure. This is the common experimental setup for all the sensing techniques tested. In general, the sensing techniques rely on measurements before and after a perturbation to the cavity. In Section VI, results on three different classes of perturbations are presented; these are: i) static boundary condition perturbations (i.e., insertion of an object) at six specified locations in the cavity, ii) perturbation of the medium of wave propagation in the cavity, and iii) global perturbation to the cavity. Next, the peculiarities of each sensing technique is discussed, and an indicator value of perturbation is defined for each sensing technique.

### IV. 1) Sensing based on "propagation comparison"

The sensing techniques that rely on "propagation comparison" work as follows. The first step is to broadcast a short pulse of a carrier signal into the cavity [See Figure 2(a)]. In the experiment discussed here, an acoustic pulse with a carrier frequency of 7kHz and a Gaussian envelope with time width of 1ms is broadcast into the stairwell. A typical input signal is shown in Figure 3(a).The carrier wave has a wavelength that is much smaller than the typical size of the cavity so that the semiclassical limit applies. The time duration and envelope of the pulse are chosen to keep the bandwidth of the pulse narrow enough to minimize the additive background noise in the cavity, which cannot be mitigated by simple band pass filtering. A center frequency and bandwidth of the pulse which result in a relatively strong coupling of the pulse energy into the cavity are chosen.

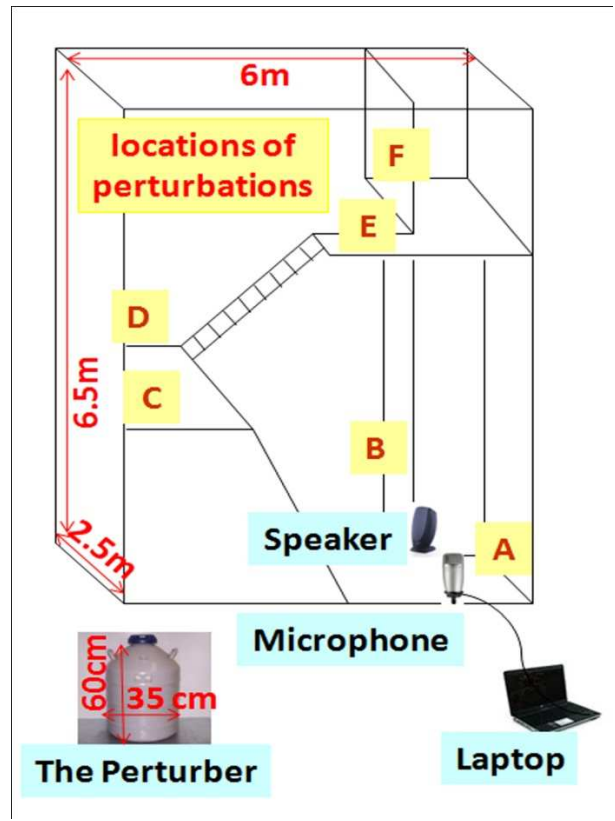

**Figure 1: The experiment is conducted inside a stairwell with cinderblock walls and tile floors. The locations of perturbations chosen to exemplify short, medium and long range detection attempts, both at concealed and non-concealed locations, with respect to the sensor, are labeled with letters A to F. The inset shows the perturbing object that is introduced at the various locations A-F.**

The second step of these sensing techniques is recording the response of the cavity to the stimulus pulse; this response is called the *sona* signal. Figure 3(b) shows a typical sona signal from the stairwell. The sona is band pass filtered using a pass-band that matches the bandwidth of the OP. The sona effectively contains multiple reflections of the pulse off different parts of the stairwell and extends in time for many pulse durations. A baseline sona signal is recorded by the microphone before perturbing the cavity [See Figure 2(b)]. For the case of 'boundary condition perturbation', the stairwell is perturbed

Figure 1. Then, the pulse is rebroadcast into the perturbed stairwell [See Figure 2(c)], and the resulting perturbed sona is recorded by the microphone [See Figure 2(d)]. The baseline sona, which is collected before the perturbation, and the perturbed sona, which is collected after the perturbation, are compared in one of the following ways giving rise to two sub-classes of sensing techniques by " propagation comparison". We refer to these techniques as Sensing by Cross Correlation (SCC) and Sensing by Mutual Information (SMI).

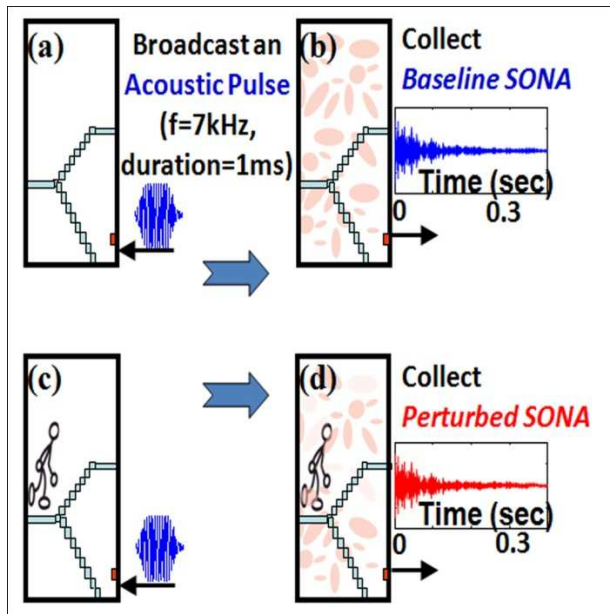

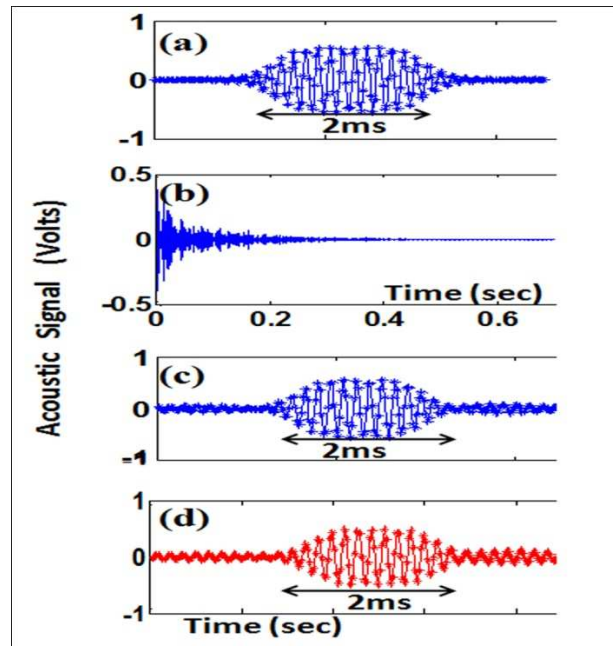

Figure 2: Schematic Operation of a sensor based on "propagation comparison". An acoustic pulse is broadcast into the stairwell in *a* and *c*. The resulting sona signals are recorded in *b* and *d*. In *c* and *d*, the cavity is perturbed. The sensor works by comparing the baseline and perturbed sonas through either cross correlation or mutual information. The red rectangle, which is at the bottom right corner of the schematic of the stairwell, schematically shows the speaker and microphone.

by inserting a cylindrical perturbing object, which has just about 0.1% of the total volume of the stairwell. The perturber, which is shown as an inset in Figure 1, is placed at one of the six perturbation locations labeled A through F in

Figure 3: a) The Original Pulse (OP) broadcast into the stairwell, b) sona c) a baseline time reversed reconstructed pulse (BRP) d) a perturbed time reversed reconstructed pulse (PRP). All parts show an acoustic signal (in Volts) versus time.

### IV.1.1) Sensing by Cross Correlation (SCC)

One way of comparing the sonas before and after perturbation involves computing the maximum of their cross correlation. As can be seen in Eqs. (1) and (2) below, this approach is inspired by the SF [23-27]. Consider two time-domain sona signals that are represented as vectors, *X* and *Y*, of voltage sample values that can be indexed in time. The cross correlation, *(X\*Y)[n]* (Eq. (1)), of these two signals is computed by finding their magnitude-normalized dot product while applying an index shift, *n*, between the signals; the cross correlation is a function of the index shift applied between the signals.

$$(X * Y)[n] = \frac{\sum_{m=1}^{m=l} X[m]Y[m+n]}{\|X\|\|Y\|}$$

**(1)**

Here, the numerator of the right hand side represents the dot product between the sona vectors *X* and *Y*, whose contents are shifted by index *n* with respect to each other; for a given value of *n*, *l* is the maximum index in which both *X[l]* and *Y[l+n]* have a well defined value. The denominator represents the product of the magnitudes of the sona vectors *X* and *Y*. The maximum of the cross correlation values (taken over all possible index shifts, *n*) is used as an indicator value of perturbation, $I_{SCC}$, for the sensing technique SCC;

$$I_{SCC} = Maximum_n\{(X * Y)[n]\}$$

**(2)**

If there is no perturbation in the cavity, the indicator value of perturbation for SCC ($I_{SCC}$) is expected to be 1; otherwise $I_{SCC}$ is generally a number between 0 and 1.

The reason for applying an index shift between the sonas while computing their normalized dot product, and later considering the maximum of the cross correlation, is as follows. The sona signals measured before and after the perturbation are digitized using slightly different time bases. In general, this is due to variations in data acquisition triggering. Thus, it is essential to align the sona signals by applying an appropriate relative index/time shift between them before considering the resulting correlation value; typically, a relative time shift of at most 20ms (i.e. an index shift of at most 880) is applied between the sonas.

**IV.1.2) Sensing by Mutual Information (SMI)**

An alternative method of comparing the two sona signals is to measure their mutual information. In the context of this computation, each sona is considered as a random variable, *X*, taking on different voltage values as time increases. A histogram of the voltage values of a sona can be constructed using equally spaced bins. The size of these bins in Volts is determined by the inherent voltage fluctuations due to measurement noise. For this experiment, different bin sizes were tried and 1mV (which is also the measurement noise level) is chosen as it resulted in an optimal detection capability of the SMI technique. Thus, slightly different voltage values of the sona, which are all within an interval whose width is the typical noise level, are considered as a single voltage value for the purpose of construction of the histogram. The probability mass function, *p(x)*, of the sona is readily derived from the histogram constructed; *p(x)* represents the probability that sona *X* has a voltage value of *x*. The entropy of the sona signal, which quantifies the information content of the sona in bits, is denoted as *H(X)*.

$$H(X) = - \sum_{x \in X} p(x) \log_2 p(x)$$

**(3)**

All the voltage values that the sona could take on after the binning process are considered in this formula for the entropy.

The mutual information of sonas *X* and *Y*, which are considered as random variables, is denoted by *I(X;Y)*, and serves as the indicator value of perturbation, $I_{SMI}$, for the SMI technique.

$$I_{SMI} = I(X;Y) = \sum_{x \in X} \sum_{y \in Y} p(x,y) \log_2 \left(\frac{p(x,y)}{p(x)p(y)}\right)$$

**(4)**

The mutual information can be described as the difference between the sum of the individual entropies of the sonas and their joint entropy *I(X;Y)=H(X)+H(Y)-H(X,Y)*. The calculation is similar to that of the entropy except that now the joint probability mass function of sonas *X* and *Y*, *p(x,y)*, is involved (Eq. 4); the marginal probability mass functions of *X* and *Y* are denoted by *p(x)* and *p(y)*. The joint probability mass function *p(x,y)* assigns the probability that sona *X* and *Y* take on voltage values *x* and *y* respectively at the same time. Once again, the bins have a size on the order of the noise level in the data.

As discussed in Section IV.1.2, the sona signals *X* and *Y*, which are collected under slightly different time bases, are time aligned based on their maximum correlation value before their correlation is considered as an indicator value of perturbation. By the same token, the computation of the joint probability *p(x,y)* of event (*X=x,Y=y*), which is used in determining the mutual information (Eq. 4) of two sona signals *X* and *Y*, is done after the sonas are aligned with respect to their time index. The alignment can be achieved by finding a time index shift between the sona signals which maximizes their mutual information.

The mutual information is zero if the two signals being compared are statistically independent. In general, the mutual information takes on values ranging from zero to a maximum value, which is the entropy value of a sona signal in the case of two identical sonas. A typical sona signal in these experiments has an entropy of about 5 bits; whereas, the mutual information between two sonas collected from two nominally identical configurations of the stairwell is typically about 2 bits.

## IV.2) Sensing based on time reversed wave propagation

The extension of the LE to classical waves is tested by using a one channel time reversal mirror for acoustic waves in the same stairwell. As in the experiment discussed above, an acoustic pulse with 7kHz center frequency and a Gaussian envelope of 1ms time width is broadcast into the stairwell. [See Figure 4(a)] The resulting sona is measured by the microphone and digitized as shown in Figure 4(b). The digitized and band-pass filtered sona is time reversed before it is broadcast back into the stairwell through the speaker. [See Figure 4(c)] To carry out a full and complete time-reversed wave propagation process, the time reversed sona should be broadcast back into the stairwell from the location of the microphone, where the sona was collected. However, spatial reciprocity of the wave equation is employed which allows us to broadcast the time reversed sona from the speaker at its original location without the need to interchange the location of the two transducers. The time reversed sona propagates in the cavity and reconstructs as a time reversed pulse at the location of the microphone, where it is recorded. [See Figure 4(d) and Figure 3(c)] The time reversed pulse is periodically generated using the same time reversed sona signal and possibly different conditions of the cavity monitored. If a perturbation occurs [See Figure 4(e)], the resulting reconstructed time reversed pulse shown in Figure 4(f) will be different from the reconstructed pulse shown in Figure 4(d).

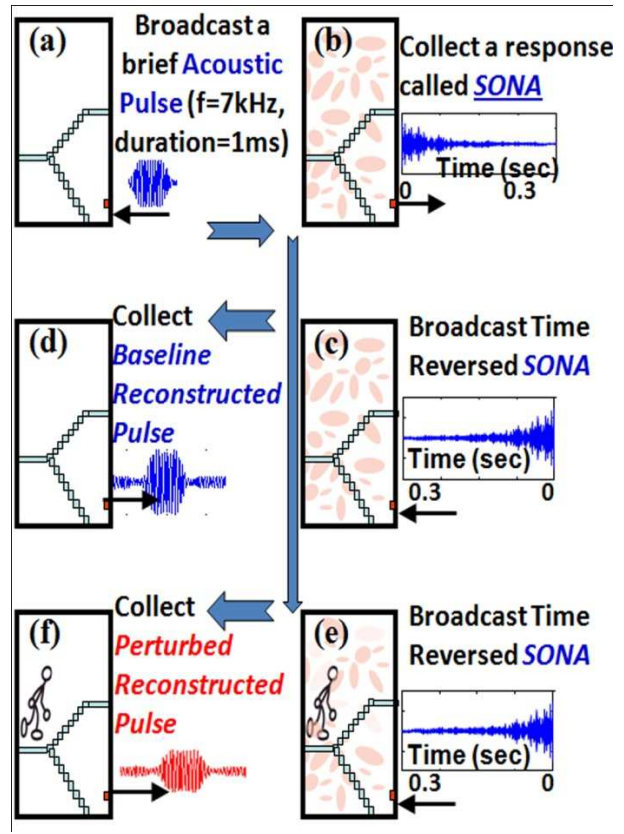

**Figure 4: Schematic Operation of the Chaotic Time Reversal Sensor, which is based on the extension of the LE to classical waves. A sequence of steps illustrated in *a,b,c,&d* are carried out to measure the Baseline Reconstructed Pulse(BRP). Using the sona collected in *b*, the steps illustrated in *e & f* are carried out to measure the Perturbed Reconstructed Pulse(PRP). The CTRS works by comparing the baseline and perturbed pulses collected.**

In general, the sensing techniques based on time reversal work by comparing two time reversed pulses reconstructed under baseline and perturbed conditions of the cavity; hence, such sensing techniques are called Chaotic Time Reversal Sensors (CTRS). The time reversed pulses reconstructed under a baseline and a perturbed condition of the cavity are referred to as a Baseline Reconstructed Pulse (BRP) and a Perturbed Reconstructed Pulse (PRP), respectively. A typical BRP and PRP are shown in Figure 3(c) and 3(d) respectively. The comparison between BRP and PRP, which have a brief time duration, is computationally inexpensive and can be done in a number of different ways. Two representative methods of comparing these signals which give rise to two versions of the CTRS, namely CTRS1 and CTRS2, are discussed. CTRS1 is based on the comparison of the peak to peak amplitude of BRP and PRP. Alternatively, CTRS2 is based on the computation of a normalized correlation of the brief pulses PRP and BRP with a time reversed version of the OP, which is shown in Figure 3(a).

### IV.2.1) Chaotic Time Reversal Sensor 1 (CTRS1)

Comparison of the BRP and PRP based solely on their peak to peak amplitude is computationally the simplest and most efficient. The ratio of the peak to peak amplitudes of the PRP to BRP is defined as an indicator value of perturbation for CTRS1, $I_{CTRS1}$.

$$I_{CTRS1} = \frac{Pk-Pk\ Amplitude_{PRP}}{Pk-Pk\ Amplitude_{BRP}}$$

**(5)**

This ratio is expected to be about 1 if the perturbed condition of the cavity is the same as its baseline condition. In the case of an actual perturbation, the ratio is a number smaller than 1. The contrast in the amplitude of BRP and PRP can be seen in Figure 3(c) and 3(d).

### IV.2.2) Chaotic Time Reversal Sensor 2 (CTRS2)

An alternative method to compare BRP and PRP is based on a normalized correlation that is analogous to the definition of the LE. Consequently, this method involves the use of the OP, which is broadcast into the cavity in order to collect the sona. The OP broadcast by the speaker is measured in a separate experiment carried out in an anechoic chamber whose walls are acoustic absorbers. Figure 3(a) shows a typical measured OP. Once the OP is measured and digitized it is numerically time reversed resulting in the Reversed Original Pulse (ROP). In principle, the ROP is expected to be identical to the BRP. However, this is not the case because the one channel acoustic time reversal mirror is not perfect. The imperfections are due to the finite time recording of the sona [29], and the dissipation in the cavity [9]; the imperfection of the time reversal process due to loss is also discussed in Section II. Additive noise from the cavity within the bandwidth of the OP also plays a role in the incongruity of the ROP and BRP.

The correlation of the ROP and the BRP is used to quantify the overall limitations of the time reversal mirror. If the experiment were ideal, in the sense that the sona were recorded for an infinite amount of time in a non-dissipative and noiseless system, this correlation would be 1 for a ray chaotic system; in these experiments this correlation is roughly 80%. This correlation is used below to normalize the correlation of the ROP and the PRP. The ratio of these two correlations is the indicator value of perturbation for the CTRS2 technique, $I_{CTRS2}$.

$$I_{CTRS2} = \frac{<PRP,ROP>/\|PRP\|\|ROP\|}{<BRP,ROP>/\|BRP\|\|ROP\|}$$

**(6)**

Here, pulses PRP, ROP, and BRP are considered as vectors of voltage values that can be indexed in time. Thus, the numerator of $I_{CTRS2}$ is the dot product of PRP and ROP divided by the product of their magnitudes. Likewise, the denominator of $I_{CTRS2}$ is the dot product of BRP and ROP divided by the product of their magnitudes. Note that the quantity in the numerator of this $I_{CTRS2}$ is analogous to the definition of the LE. The normalization in the denominator is needed to ensure that $I_{CTRS2}$ is 1 in the absence of a perturbation. In the presence of a perturbation the $I_{CTRS2}$ is a number between 0 and 1.

Yet another way of comparing the BRP and PRP is their correlation, $\langle BRP,PRP \rangle / \|BRP\|\|PRP\|$. However, we have experimentally demonstrated that such an approach does not yield a reliable indication of whether a perturbation has happened or not. In other words the correlation of two time reversed pulses that are reconstructed before and after perturbation is not statistically distinguishable from the correlation of two time reversed pulses that are reconstructed under nominally identical conditions of the cavity. Hence, this third variety of the CTRS is not discussed further.

### IV.3) Effects of dissipation and processing the sona signal:

The sensing techniques discussed so far face the problem of dissipation of classical waves which effectively limits the sensitivity and spatial coverage of the sensor. The dissipation brings about an exponential decay of the signal set up by the initial broadcast of the acoustic pulse. This exponential decay is seen in the envelope of the sona signal recorded, from which the *1/e* decay time is estimated. [Figure 5(a)] A typical 1/e decay time of the sona signals collected from the stairwell is about 0.1 seconds. This measured 1/e decay time, $\tau$, is reasonably consistent with the 60 dB decay time of the stairwell estimated from Sabine's formula.

$$T_{60dB} = cV / \sum_m S_m \alpha_m$$

**(7)**

Here, the parameter c= 0.161 s/m. Applying Sabine's formula involves estimating the volume of the cavity, $V \approx 93 m^3$. In addition, the surface area, $S$, of each of the constituent materials of the interior of the cavity is estimated. The corresponding frequency dependent absorption coefficient, $\alpha$, of the materials is found from the literature [30], and the summation in Eq. 7 is carried out over all the constituent materials, $m$, of the interior of the cavity. The interior of the stairwell has approximately $129 m^2$ of painted concrete block and $46 m^2$ of concrete floor; these constituent materials are known to have an absorption coefficient of 0.08 and 0.02 respectively for 4kHz sound waves. Using these rough estimates, the 60 dB decay time for 4kHz sound waves in the stairwell is 1.3s. From this, one estimates a *1/e* decay time of 0.09 seconds for 4kHz sound waves in the stairwell, which is close to the measured $\tau$ =0.1s at 7kHz.

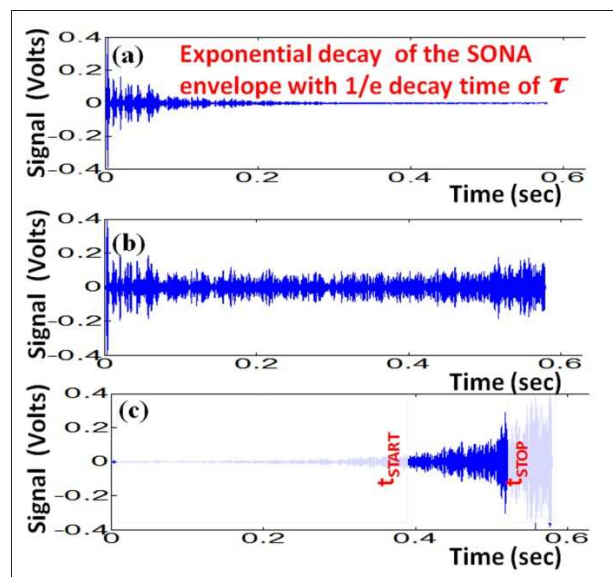

**Figure 5: a) A typical measured exponentially decaying sona signal, b) Exponentially amplified sona with F=1, c) Exponentially amplified sona with F=2 after rectangular windowing between times $t_{START}$ and $t_{STOP}$.**

The exponential decay of the sona can be numerically mitigated by applying an exponential amplification *A(t,F)* to the portion of the sona signal that has a Signal to Noise Ratio (SNR) of at least 1.

$$A(t, F) = e^{Ft/\tau}$$

**(8)**

The time dependent amplifying function, *A(t,F)*, is a function of parameter *F*, and it uses the measured value of the *1/e* decay time, *τ*, of the sona signal being amplified [31]. The parameter F typically takes on values of either 0, 1, or 2. If F=0, there is no exponential amplification of the sona. [See Figure 5(a)] If F=1, the resulting exponential amplification removes the effects of dissipation that happened during the time-forward propagation of the acoustic pulse up to the collection of the sona. [See Figure 5(b)] If F=2, the resulting exponential amplification removes the effects of dissipation that the sona has suffered up to its collection during time-forward propagation and also the dissipation that it will suffer as it goes through the stairwell again in a time reversed manner. [See Figure 5(c)]

The motivation for applying exponential amplification is to make the sona signal closer to what it would be in the non-dissipative case. Working in the approximately non-dissipative case can expand the range of the sensor. In addition, the range of the sensor can be changed to some extent with choice of parameter value *F*. In Section VI.3, we shall see that global perturbations to the stairwell are detected best when the sonas are exponentially amplified to approximate the non-dissipative case. The exponential amplification is also motivated by the theoretical results in Section II.

Another possibility of tuning the sensor involves applying a rectangular time-gating window function to the sona. Such a window is a function of two parameters: start and stop time (See Figure 5(c)). The motivation for time-windowing the sona to change the sensitivity and spatial coverage of the sensor is founded on a 'ray propagation model' of the problem. Rays that bounce back from perturbation locations in the vicinity of the sensor get recorded at the beginning of the sona. In contrast, rays that bounce back from perturbation locations farther out from the sensor are recorded towards the end of the sona. This simple generalization of the complex ray trajectory dynamics in the stairwell motivates the possibility of windowing the sona to change the spatial sensitivity of the sensor. Thus, the start and stop time parameters of the rectangular window are varied to explore this possibility of tuning the sensor's sensitivity to perturbation at various locations within the stairwell.

The rectangular time window has a rise and fall time that is designed to keep the bandwidth of the windowed sona invariant. Particularly, the rise and fall times are both on order of magnitude of the time width of the original acoustic pulse that generated the sona (i.e. 1ms). Before a sona is windowed, it is exponentially amplified with a given F value. The amplitude of the windowed sona is then uniformly scaled to fit into the voltage dynamic range of linear output of the speaker, which is -0.4V to 0.4V. This range was determined by an experiment in an anechoic chamber with the microphone and speaker.

To summarize, the sona signal is processed using the three parameters discussed above: exponent *F*, start time, and stop time. Each of the sensing techniques discussed so far are done with various values of these parameters. For sensing techniques based on time reversal of wave propagation, the sona is processed with the appropriate parameters before it is time reversed and broadcast back into the cavity. On the other hand for the sensing techniques based on "propagation comparison", both of the sonas being compared are processed by the same exact parameter values before the computation of mutual information or cross correlation of the processed sonas are carried out.

As a caveat, the following special procedures are taken for the case of the SMI technique to improve its detection performance. The windowed sonas are not uniformly scaled to fit into the dynamic range of -0.4V to 0.4V (mentioned above). Furthermore, the voltage values of the processed sonas are rounded off to 3 significant figures both before and after the sonas are processed (using exponential amplification and windowing); in other words, the binning of the sonas with a bin size of 1mV is done both before and after processing the sonas.

### IV.4) Investigation of the tunability of the range of the sensor

So far, four sensing techniques have been introduced, and a mechanism to tune the range of a sensor using three parameters is established. The parameters are designed to compensate for the effects of dissipation and to alter the spatial range of the sensor, to some extent. The following experiments were done to investigate the problem of perturbation detection at short, medium and long range.

Six different locations of perturbations, which are labeled A through F in Figure 1, were chosen in the stairwell. These locations were chosen so that there are two representative locations for short (perturbation locations A and B), medium (perturbation locations C and D) and long range (perturbation locations E and F) detection attempts, respectively. Each pair of representative locations were chosen so that there is an example of a location that is concealed from the sensor (B, D, and F), and a location that is almost within the line of sight of the sensor, or at least within a couple of reflections from the sensor (A, C, and E). For each sensing technique, the baseline (unperturbed) situation involves the absence of the perturber in the stairwell, while the perturbed situation has the perturbing object located at one of the six locations A through F.

The detection experiment was systematically performed at each perturbation location using all the sensing techniques introduced above. The experiment was carefully designed to allow all the sensing techniques to be applied to a single instance of perturbation at a given location. All the sensing techniques were operated with the same set of parameter values. This experimental scheme allows for the following considerations. An optimal set of parameter values can be identified for a given sensing technique at a given perturbation location. The effectiveness of a sensing technique, which is operating at its optimal parameter values, can be gauged at different perturbation locations. The optimal detection capability of different sensing techniques can be compared at a given perturbation location. Standardization of these comparisons is discussed in the data analysis section.

### V. DATA ANALYSIS

In the experiment section, the measurement and calculation of four different indicator values of perturbation (i.e. $I_{CTRS1}$, $I_{CTRS2}$, $I_{SCC}$, and $I_{SMI}$) corresponding to the four sensing techniques were introduced. Each of those indicator values have their own inherent uncertainty in their measurement and calculation. The range of values that the indicators take on is not uniform. Even though $I_{CTRS1}$, $I_{CTRS2}$, and $I_{SCC}$ have the same range of values (i.e. 0 to 1), the dependence of their value on the perturbation is not necessarily the same. All these complications make the comparison of the different sensing techniques, solely using their respective indicator values of perturbation, a difficult task. This problem is solved by defining a standardized Figure of Merit (FOM) that can be calculated from the typical statistics of the indicator values of any of the techniques.

In the absence of perturbation, $I_{CTRS1}$, $I_{CTRS2}$, and $I_{SCC}$ should ideally be 1. Whereas, $I_{SMI}$ should have a particular value, which is closest to the typical entropy of the sona in bits, in the absence of perturbation. However, this is not always the case due to measurement uncertainties and noise that propagate through the steps of the computation of the indicators. Consider a control experiment of detection, in which we do not induce any perturbation to the cavity under surveillance. In such a control experiment, the resulting indicator values fluctuate somewhere around the ideally expected value of 1 (for CTRS1, CTRS2 and SCC), or somewhere around a value close to the entropy of the sona in bits (for SMI). The statistics of these control indicator values of perturbation are considered for each sensing technique. Particularly, the mean, $\mu$, and standard deviation, $\sigma$, of the control indicator values of perturbation are calculated.

If an indicator value of perturbation is much smaller than the mean of the control indicator values compared to their standard deviation, then there is a statistically significant detection. Thus, the following Figure of Merit(*FOM*) is defined.

$$FOM = \frac{\mu - I}{\sigma}$$

(9)

The *FOM*, is the ratio of the difference between the observed indicator value, *I*, and the mean, $\mu$, of the control indicator values to the standard deviation, $\sigma$, of the control values. The observed indicator value of a perturbation, *I*, for a given instance of perturbation may itself fluctuate around some value due to noise. This results in the *FOM* fluctuating as well. Therefore, the *FOM* is averaged over 25 different realizations. Such an average *FOM*, *<FOM>*, also has a propagated uncertainty, $\delta_{<FOM>}$, associated with it. The difference between the average *FOM* and the uncertainty in the average *FOM* is defined as the Lower-bound of the Figure of Merit, *FOM$_L$*.

$$FOM_L = \langle FOM \rangle - \delta_{<FOM>}$$

(10)

To use an abundance of caution, the *FOM$_L$* is used to ultimately decide whether or not there is a statistically reliable detection. Heuristically, if *FOM$_L$* is greater than 2, then we conclude that there is a statistically reliable detection.

The *FOM$_L$* is calculated for detection attempts using different parameter values. In what follows, the *FOM$_L$* is plotted, using a contour plot, as a function of start time and stop time parameters of the rectangular time-windowing function applied to the sona. [See Figure 6]These plots are done for a given value of the *F* parameter used to amplify the sona. Such plots are also annotated by the sensing technique that was used to generate the *FOM$_L$* and also the location of the perturbation that is being detected.

## VI. RESULTS

### VI.1) Results of detection of perturbations at specified locations

The experiments performed can be summarized as follows. Detection attempts were made using four different sensing techniques at six different perturbation locations in the stairwell, which are labeled A through F in Figure 1. The six perturbation locations are chosen to be representative of short, medium and long range detection both in a concealed and non-concealed sections of the stairwell with respect to the sensor. Each of the detection attempts using each technique were done using various parameter values. Particularly, the *F* parameter, which controls the exponential amplification, takes on values of 0, 1 or 2. The start time and stop time of the windowing function each take on 7 equally spaced values ranging from 0 seconds to the time at which a typical sona's SNR becomes 1, which is roughly 0.7 seconds. Therefore, there are (7*(7-1))/2=21 plausible pairs of start time and stop time values that constitute a rectangular sona windowing function of non-zero time width.

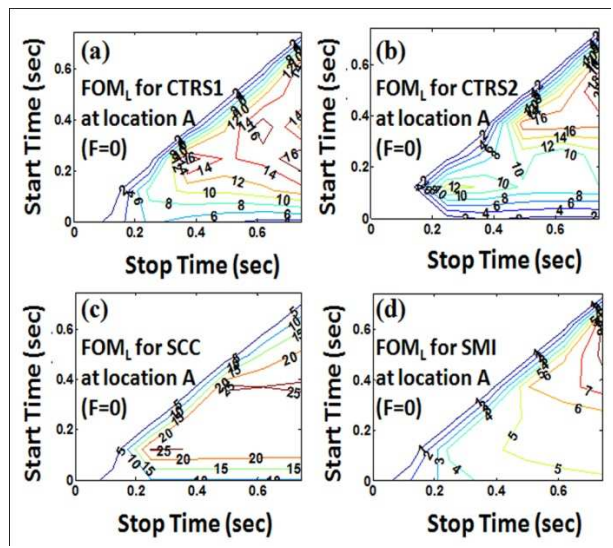

**Figure 6: Contour plots of the lower bound on the Figure of Merit(FOM$_L$) as a function of start time and stop time parameters of the rectangular time windowing function applied to the sona. The plots show detection attempts at perturbation location A (indicated in Figure 1) using F=0. a) FOM$_L$ for CTRS1, b) FOM$_L$ for CTRS2, c) FOM$_L$ for SCC d) FOM$_L$ for SMI.**

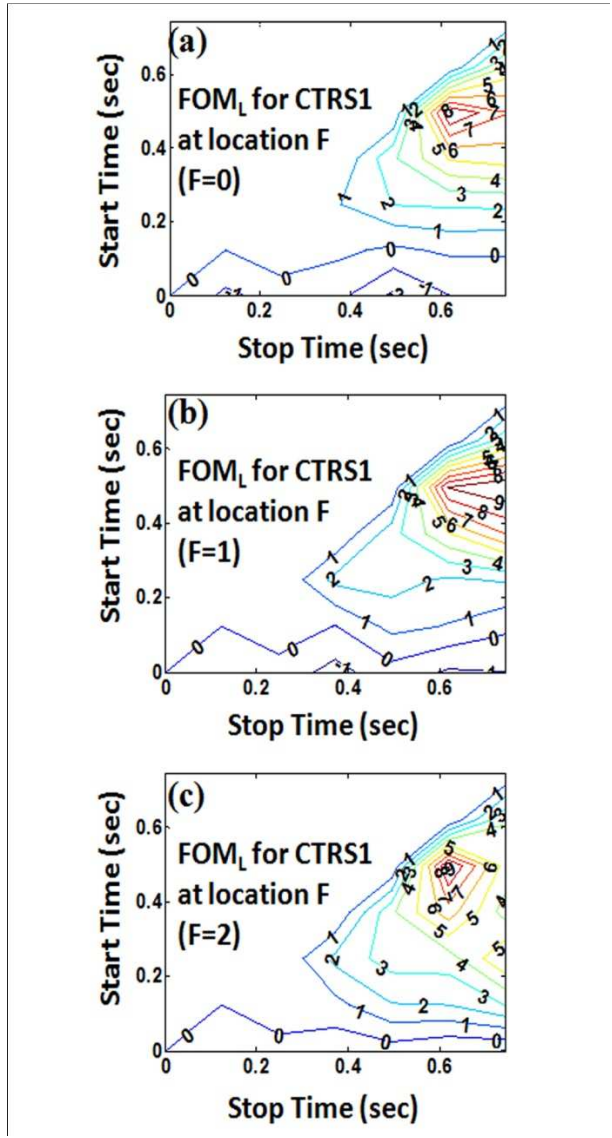

**Figure 7: Contour plots of the Lower bound on the Figure of Merit (FOM$_L$) as a function of start time and stop time parameters of the rectangular windowing function applied to the sona. a) long range detection at location F, indicated in Figure 1, using CTRS1 with F=0, b) long range detection at location F using CTRS1 with F=1, c) long range detection at location F using CTRS1 with F=2.**

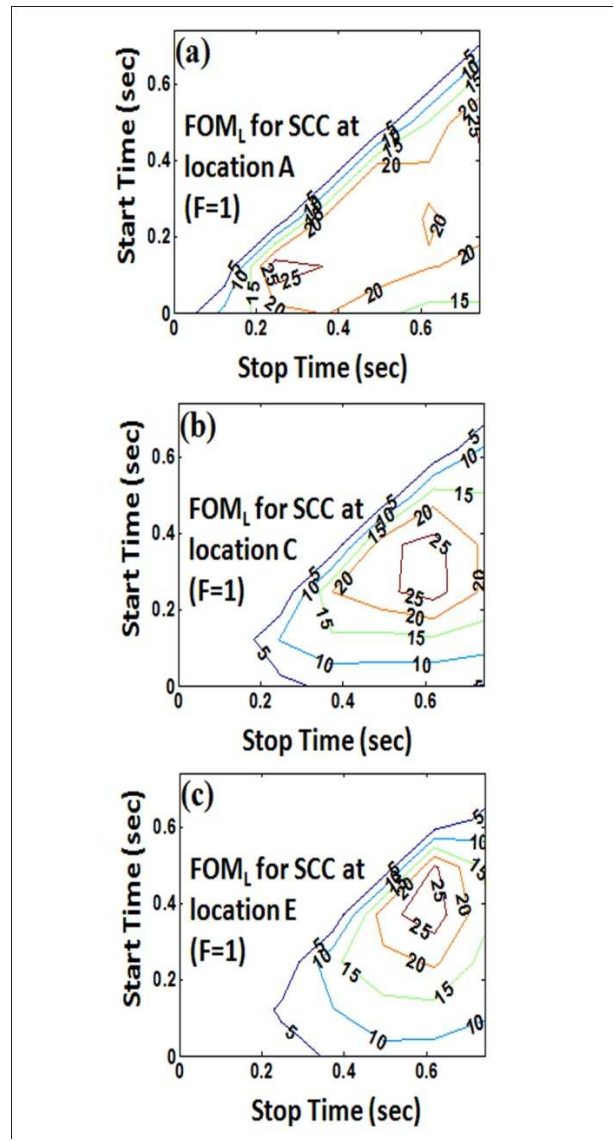

**Figure 8: Contour plots of the Lower bound on the Figure of Merit(FOM$_L$) as a function of start time and stop time parameters of the rectangular windowing function applied to the sona. In all plots, an exponential amplification of F=1 is applied to the sona. The three plots shown here show detection attempts at different locations of perturbations illustrated in Figure 1. a) short range detection at location A, b) medium range detection at location C, c) long range detection at location E.**

The *FOM$_L$* (Eq. 10), which is a function of start time and stop time, is plotted as a contour plot for a specified *F* value, sensing technique and perturbation location. In such contour plots, only the lower right triangle of the plane is used.

| Sensing Technique | | Perturbation Location as indicated in Figure 1 | | | | | |
|---|---|---|---|---|---|---|---|
| | | A | B | C | D | E | F |
| CTRS1 | Maximum $FOM_L$ value | **28.9** | **24.6** | **17.8** | **16.0** | **20.1** | **9.9** |
| | % of $FOM_L$ value >2 | **97%** | **81%** | **73%** | **71%** | **67%** | **46%** |
| CTRS2 | Maximum $FOM_L$ value | **21.7** | **18.0** | **9.9** | **11.5** | **13.1** | **5.8** |
| | % of $FOM_L$ value >2 | **90%** | **63%** | **67%** | **32%** | **48%** | **13%** |
| SCC | Maximum $FOM_L$ value | **32.7** | **33.1** | **33.1** | **22.3** | **30.2** | **7.9** |
| | % of $FOM_L$ value >2 | **100%** | **95%** | **87%** | **78%** | **87%** | **24%** |
| SMI | Maximum $FOM_L$ value | **8.3** | **11.2** | **12** | **8.3** | **15.2** | **3.5** |
| | % of $FOM_L$ value >2 | **76%** | **71%** | **89%** | **54%** | **76%** | **6%** |

**Table I: The maximum $FOM_L$ over all the parameter values tried is shown for each of the four sensing techniques detecting a perturbation at each of the six perturbation locations indicated in Figure 1. In addition, the percentage of parameter values which gave a $FOM_L$ that is greater than 2 is also shown.**

Overall, since there are 6 perturbation locations, 4 sensing techniques and 3 F-values, there are 72 such contour plots for the set of experiments carried out. In this results section, a select group of these plots, which illustrate general trends, will be presented. A table that summarizes all the results is also included. [See Table (I)] Given a perturbation location and sensing technique, the table shows the maximum $FOM_L$ value over all parameter values tried in these experiments. The table also shows the percentage of parameter values that gave a $FOM_L$ greater than 2, which is a conservative estimate of statistically reliable detection. The table gives an overall sense of the effectiveness of the sensing techniques, because it presents their performance in detecting perturbations at different ranges from the sensor.

In Figure 6, the $FOM_L$ is plotted for detection attempts at perturbation location A (shown in Figure 1) without exponential amplification of the sona (i.e. F=0). Figures 6(a), (b), (c), and (d) demonstrate that the techniques of CTRS1, CTRS2, SCC, and SMI, respectively, allow for a short range and non-concealed perturbation detection over a wide range of parameter values (i.e. $FOM_L$ is greater than 2 for a large number of rectangular windowing functions). The SMI technique has relatively smaller $FOM_L$ values compared to the other techniques. Overall, all the sensing techniques work without the need for exponential amplification and windowing of the sona when the perturbation is in the vicinity of the sensor.

Here, Figure 6c illustrates the connection between the calculations of the SCC technique and the traditional SF [23-27], which has inspired the SCC technique. In Figure 6c, there is no exponential amplification (i.e. F=0). Therefore, the $FOM_L$ values plotted near the diagonal-line of the start-time stop-time contour plane essentially come from a set of $I_{SCC}$ values which can be plotted as SF versus time of the baseline and perturbed sona signals being compared. We see that the optimal parameter region in Figure 6c is not near the diagonal-line of the contour plane; thus, the generalized SCC technique does indeed offer greater flexibility with its three adjustable parameters (start-time, stop-time, & F), especially for perturbations that are further from the sensor, and/or hidden.

The need to process the sona comes into play when a medium or long range detection is attempted. In Figure 7, the results of long range detection at concealed perturbation location F [See Figure 1] are presented. The $FOM_L$ for the CTRS1 technique is plotted with $F$=0, $F$=1, and $F$=2 in Figure 7(a), (b), and (c) respectively. In contrast to Figure 6(a) (which shows results for short range detection by CTRS1 with $F$=0), a smaller set of windowing parameters allow long range detection by CTRS1 with $F$=0. Therefore, successful long range detection demands a judicious choice of windowing parameters with $F$=0. If there is an exponential amplification with $F$=1 or $F$=2, there is, in this case, a slightly larger set of windowing parameters that can be used to do long range detection. However, as can be seen in Figure 7, the right choice of the windowing parameters is more important in doing long range and concealed detection using CTRS1 than the value of $F$; this is also generally true for the CTRS2 and SCC. In general, the percentage of parameter values that allow detection decreases as the perturbation location gets farther away from the sensor, as seen in Table (I).

The possibility of associating a set of optimal detection parameter values with detection of a perturbation at a particular location was investigated next. In general, as the perturbation location is farther away from the sensor, the optimal detection parameters space either shrinks and /or moves to the upper right corner of the "start time - stop time plane". Figure 8 illustrates this phenomena for the case of short, medium and long range detection attempts at perturbation locations A, C and E [See Figure 1] respectively by the SCC. The broad swath of parameter space that is optimal for detection at short range [See Figure 8(a)] shrinks as the perturbation location moves farther away from the sensor [See Figures 8(b) and 8(c)]; it also moves to the upper right corner of the plane in this case. Even though this is a consequence of the fact that the waves that bounced off the farthest perturbation location take a longer time to get back to the sensor, it is not a trivial consequence as there are multiple reflections of all the waves within the cavity.

### VI.2) Results on detection of perturbations of the medium of wave propagation in the cavity

So far, the results of experiments which involve detection of perturbations at six different locations in the stairwell, illustrated in Figure 1, are presented. Such perturbations essentially change the boundary conditions of the cavity at a localized region. A different kind of perturbation involves perturbation of the medium of wave propagation in the cavity: For example, creating air currents will perturb acoustic wave propagation. Such perturbations naturally start out locally and may spread out throughout the medium filling the cavity in a complex manner. This motivates yet another kind of perturbation to the cavity which is global in nature. As a significant amount of time elapses, both the boundaries of the cavity and the medium within may undergo complex and spatially extensive changes due to uncontrollable thermal variations (giving rise to convection currents, for example). Next, we present the results of experiments which investigate the possibility of detecting a relatively localized perturbation to the medium of wave propagation in the cavity, and also a global perturbation to the cavity.

The medium of wave propagation in the stairwell is perturbed by remotely activating a fan which is stationed inside the stairwell about 2m away from the sensor. The air currents induce a phase shift, $\Delta\phi$, in the sound waves that pass through the part of the cavity in which the air is perturbed.

$$\Delta\phi \approx \frac{\Delta v}{v} k L_{path}$$

(11)

Here, $v$ is the speed of sound, $\Delta v$ is the speed of the wind, $k$ is the wave number of the sound wave, and $L_{path}$ is a typical path length of travel of the sound wave through the moving air. Taking $v$=343 m/s, $\Delta v$=2 m/s, $k$=2π/5cm, $L_{path}$=1m, gives $\Delta\phi$=0.23 π. Such a significant phase shift degrades the reconstruction of the time reversed pulse during the operation of the CTRS. This is because the coherent superposition of the time reversed sona is thwarted due to the phase shift that waves, which pass through the moving air, experience.

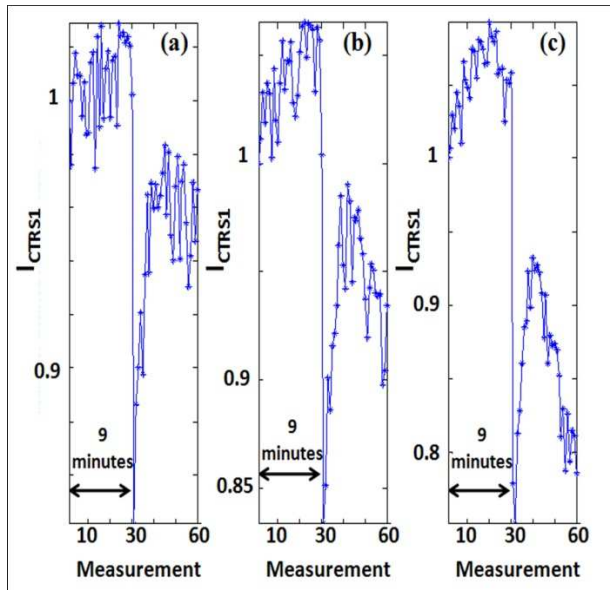

**Figure 9: Indicator values of perturbation for CTRS1, $I_{CTRS1}$, versus measurement number (approximately 18 seconds elapse between each measurement). Halfway in the displayed time interval, a mechanical fan is briefly activated in the stairwell perturbing the medium of wave propagation. Each of the plots correspond to cases in which the sona is exponentially amplified by different $F$ values. In all cases the sonas are windowed with start time=0s and stop time=0.3s: a) $F$=0, b)$F$=1, c)$F$=2.**

The following experiment is done to study the detectability of perturbations of the medium of wave propagation by CTRS1. A pulse is broadcast into a quiescent stairwell, and a sona is collected. The time reversed sona is then periodically broadcast into the stairwell 30 times over 9 minutes. The resulting time reversed reconstructed pulses are saved. Then, the air in the cavity is perturbed by remotely activating a mechanical fan for 15 seconds; the fan had been stationed inside the stairwell in the vicinity of the sensor. After the fan is turned off, the time reversed sona is broadcast into the stairwell 30 more times over 9 minutes. The resulting 30 additional reconstructed pulses are also saved. In this experiment, the very first time reversed reconstructed pulse is considered as a Baseline Reconstructed Pulse (BRP). All the other pulses are considered as a Perturbed Reconstructed Pulse (PRP). Then, the indicator value of perturbation for CTRS1, $I_{CTRS1}$, (Eq. 5) is constructed for each of the 59 PRP, BRP pairings. Finally, $I_{CTRS1}$ is plotted versus time as shown in Figure 9.

Figures 9(a), (b), and (c) show the cases when the sona is exponentially amplified with parameter $F$=0, $F$=1, and $F$=2, respectively. In all cases, the sonas are windowed with start time=0s and stop time=0.3s. From Figure 9, it is clear when the medium perturbation occurred (i.e. halfway in the displayed time axis between index 30 and 31). The dynamic nature of the perturbation is exhibited in the plots because the $I_{CTRS1}$ increases as the air currents damp out and the perturbation in the vicinity of the sensor relaxes. In Figure 9(a) there is no exponential amplification, hence the dynamic perturbation is no longer sensed after about 3 minutes, which is roughly the time that it takes for the air in the vicinity of the sensor to calm down. The $I_{CTRS1}$ indicator ends up with a smaller static value after 3 minutes in Figure 9(a) in part because after the fan is activated its blades took on a different position, which by itself is a static perturbation. However, if there is exponential amplification, $I_{CTRS1}$ changes non-monotonically as shown in Figure 9(b) and 9(c), because the sensor is now sensitive to what happens farther out, both from the fan and the sensor. In other words, the medium of wave propagation perturbation eventually spreads out in the cavity initiating a more global perturbation. In the next sub-section, global perturbations are studied in detail.

To summarize, the general results presented in this sub-section based on the CTRS1 technique are also observed in the other three techniques. It is also important to note the practical implication of these results. The medium of wave propagation can be perturbed in a variety of circumstances of interest. For instance: the dynamic nature of these perturbations means that one can verify that a cavity had been perturbed by a fast moving object even after the object has left the cavity, based solely on the air turbulence the fast moving object induced.

| Sensing Technique | | Global Perturbation | | |
|---|---|---|---|---|
| | | *F=0* | *F=1* | *F=2* |
| **CTRS1** | Maximum $FOM_L$ value | **12.5** | **13.0** | **33.0** |
| | % of $FOM_L$ value >2 | **100%** | **95%** | **90%** |
| **CTRS2** | Maximum $FOM_L$ value | **15.2** | **12.3** | **23.0** |
| | % of $FOM_L$ value >2 | **100%** | **95%** | **90%** |
| **SCC** | Maximum $FOM_L$ value | **40.3** | **56.8** | **40.6** |
| | % of $FOM_L$ value >2 | **100%** | **100%** | **100%** |
| **SMI** | Maximum $FOM_L$ value | **13.8** | **5.1** | **3.1** |
| | % of $FOM_L$ value >2 | **95%** | **90%** | **38%** |

**Table II: The maximum *FOM$_L$* over all the windowing parameter values tried is shown for each of the four sensing techniques detecting a global perturbation with a given value of the exponential amplification parameter *F*. In addition, the corresponding percentage of windowing parameter values which gave a *FOM$_L$* that is greater than 2 is also shown.**

### VI.3) Results on detection of global perturbations to the cavity

Experimentally inducing a uniform global perturbation to a cavity is not simple. A possible global perturbation is to allow the boundaries of the stairwell and its medium to undergo thermal changes through time. If sufficient time elapses, three sides of the stairwell are exposed to the outside environment, and undergo some thermal changes that approximate global perturbations.

The procedure of the global perturbation experiment in the stairwell is very similar to the procedure of the experiments performed to detect perturbations at the six locations illustrated in Figure 1. The perturbation simply involves allowing about 2 hours to elapse in between collection of baseline and perturbed sonas (time reversed pulses).

First, the same set of parameter values and techniques are used to analyze this global perturbation as in the case of the six local perturbations discussed in Section VI.1. The results are summarized in Table (II). The Table presents results for each of the 3 different exponential amplification parameter *F* values used (0, 1, and 2) separately.

From Table (II), it is seen that global perturbations can be detected by almost any of the windowing parameters tried. This supports the intuition that the effect of global perturbations leaves a signature throughout the sona signals. This raises the following question. Are global perturbations detected best when exponential amplification is applied to approximate the non-dissipative case?

If the answer is yes, then it is expected that *F*=1 is optimum for the SCC technique and *F*=2 is optimum for the CTRS1 and CTRS2 techniques. This hypothesis is tested by repeating the experiment discussed above using a different set of parameter values. Here, *F* values ranging from 0 to 3 with increment of 0.1 are used (as opposed to using just *F*=0, 1, and 2). On the other hand, no windowing is applied to the sona to simplify the experiment. As shown in Figure 10, the FOM$_L$ has a maximum around *F*=1 and *F*=2 for SCC and CTRS2 (and also CTRS1) techniques, respectively. Therefore, global perturbations are best detected by CTRS and SCC when the sona is exponentially amplified to approximate the non-dissipative case.

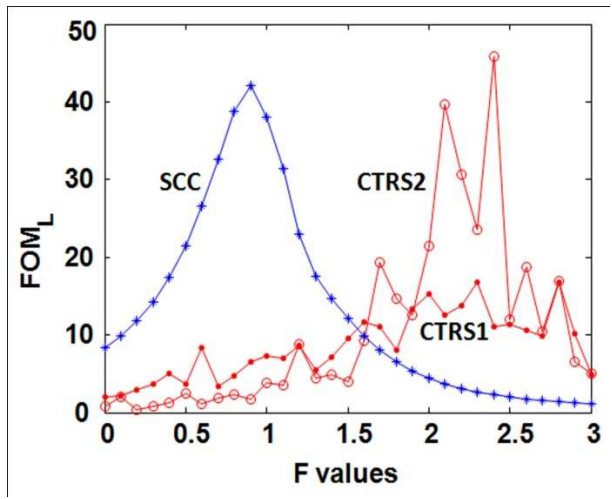

**Figure 10:** The Lower bound on the Figure of Merit (*FOM$_L$*) versus the exponential amplification parameter *F* for detection of global perturbation using CTRS1, CTRS2, and SCC. The CTRS based techniques work best when *F* is close to 2, and SCC works best when *F* is close to 1.

## VII. DISCUSSION

The results summarized in Table (I) indicate that SCC, CTRS1 and CTRS2 perform reliably in detecting perturbations at different ranges. The SMI performs detection as well, despite its relative weakness. The SCC has the highest *FOM$_L$* across the board, which is its main advantage. However, the SCC also has the broadest optimal parameter space, which may be a disadvantage if one is interested in associating a given perturbation location with a narrow distinct optimal parameter space; such an association can be useful to localize the perturbation.

Another shortcoming of the SCC, and also the SMI, is their higher computational cost. The lower bound on the computational resources needed to compare two sona signals using the SCC or SMI roughly scales with the length of the sona signals. [See Eqs. 1-4] Besides, it is important to note that we have implemented the SMI by calculating the mutual information with the so called "equidistant binning estimator" technique which is the simplest method computationally [32]; if the SMI were to be implemented using other more complicated "mutual information estimators", its higher computational cost would overshadow any other benefits. This computational problem inherent in the SCC and SMI methods can be mitigated only by considering narrow windows of the sona signals. In contrast, the CTRS based sensing techniques have a small fixed computational cost in comparing the time reversed pulses regardless of the values of the parameters used to process the sona. Particularly, CTRS1 is the most computationally efficient sensing technique as it relies on a simple peak to peak amplitude measurement of the reconstructed time reversed pulses. However, the CTRS requires analogizing and broadcasting a time reversed sona signal.

It is worth emphasizing that the SCC technique is motivated by the SF [23-27]. As mentioned in Section IV.1, the SCC technique implicitly calculates the SF for the case of no exponential amplification (i.e. F=0) as long as a set of start-time and stop-time windowing parameters are used which effectively result in the sliding of a narrow rectangular-time-window across the baseline & perturbed sona signals being compared. Figure 11 shows the 25-realization-averaged SF versus time measured for the six local perturbations illustrated in Figure 1. The SF is simply calculated using Eqs. (1) and (2) as follows: $SF(t^*)=I_{SCC}(t^*)$, where t* is the middle of the time-window formed by the parameters start-time and stop-time, and where F=0.

It is clear that the SF decays the fastest for short-range perturbations (i.e. perturbation locations A&B in Figure 1), whereas the slowest SF decay is for the long-range and concealed perturbation (i.e. perturbation location F). When the SF decay of concealed and non-concealed perturbations that are at about the same distance from the sensor is compared, it turns out that non-concealed perturbations (i.e. A, C, and E) result in a faster SF decay. These results agree with our earlier observations regarding the dependence of the optimum parameter space (in the start-time stop-time contour plane of the $FOM_L$) on the perturbation location; (i.e. long range and concealed perturbations are detected better if we look at the end of the sona.

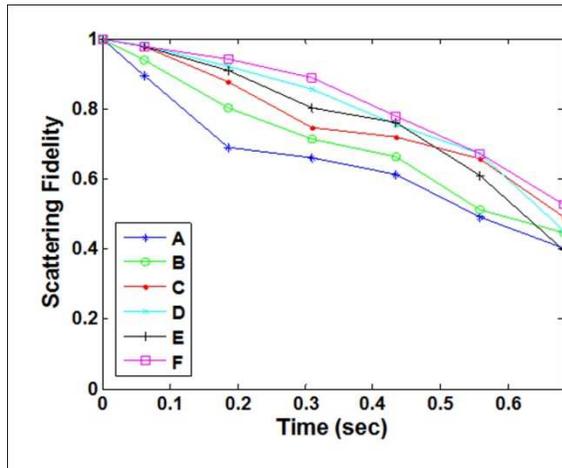

**Figure 11: SF of sonas before and after a perturbation as a function of time. The SF plotted here is averaged over 25 realizations. The width of the time window over which the $I_{SCC}$ and hence the SF is computed is 0.1 seconds. The six SF curves are labeled A through F; the labels correspond to the locations of the perturbations illustrated in Figure 1. For example, the slowest decaying SF curve comes from sonas measured before and after perturbing the cavity at location F in Figure 1. The rates of fidelity decay are generally indicative of the relative distance of the perturbations from the sensor.**

The exact mathematical equivalence between the quantum mechanical quantities LE and Quantum Fidelity is not quite replicated in their classical analogs developed here, namely the $I_{CTRS2}$ and $I_{SCC}$. It is important to note that the overlap between the classical wave systems is done only at a single point in space where the microphone is located; in addition, there is dissipation in the classical system.

## VIII. CONCLUSION

The direct analogy of the quantum mechanical concepts of LE and Quantum Fidelity give rise to the sensing techniques CTRS2 and SCC presented here. In addition, the CTRS1 and the SMI techniques are developed in parallel. The CTRS based techniques, which rely on a time reversal mirror, offer a computationally cheap alternative to the SCC and SMI techniques that are based on a more traditional "propagation comparison" concept.

A systematic set of experiments are done to detect perturbations at six different locations in an enclosed stairwell using these sensing techniques. The processing of the sona signals by exponential amplification and time windowing allowed long range detection at concealed locations in the cavity; such detection endeavors would not have been possible without such processing of the sona, especially the time windowing. The optimal parameter space of the sensing techniques is also seen to be related to the perturbation location. Even though there may not be a one to one correspondence between an optimal parameter space and a perturbation location (which would enable exact localization of the perturbation), the current results indicate that one can at least rule out candidate locations for a detected perturbation by looking at the optimal parameter space found.

In addition to detection of static boundary perturbations at given locations, perturbations to the medium of wave propagation are also shown to be detectable. Detection of such perturbations opens up a wide range of applications. It is also shown that by using exponential amplification of the sona, one can see how the initially localized medium perturbation spreads out into other parts of the cavity. The extreme case of global perturbations, which can be experimentally realized by allowing the cavity to undergo thermally induced changes, is also investigated. It is shown that the global perturbations are detected best when the sona is exponentially amplified to approximate the lossless case.

Page **21** of **21**
**ACKNOWLEDGEMENT**

This work is supported by an ONR MURI entitled "Exploiting Nonlinear Dynamics for Novel Sensor Networks" grant N000140710734, the AFOSR under grant FA95501010106, and the Maryland Center for Nanophysics and Advanced Materials. We would like to thank Michael Johnson, D. Lathrop and T.H. Seligman for helpful advice.

**REFERENCES**

[1] R.K. Snieder, J.A. Scales, *Phys. Rev. E* 58, 5668 (1998).
[2] H.-J. Stockmann, *Quantum Chaos*, Cambridge University Press, New York 1999.
[3] X. Zheng, T.M. Antonsen Jr., E. Ott, *Electromagnetics* 26, 3 (2006).
[4] X. Zheng, T.M. Antonsen Jr., E. Ott, *Electromagnetics* 26, 37 (2006).
[5] J. A. Hart, T.M. Antonsen, Jr., E. Ott, *Phys. Rev. E* 79, 016208 (2009).
[6] J.A. Hart, T.M. Antonsen, E. Ott, *Phys. Rev. E* 80, 041109 (2009).
[7] S. Hemmady, X. Zheng, E. Ott, T.M. Antonsen, S.M. Anlage, *Phys. Rev. Lett.* 94, 014102 (2005).
[8] S. Hemmady, X. Zheng, T.M. Antonsen, E. Ott, S.M. Anlage, *Phys. Rev. E* 71, 056215 (2005).
[9] B.T. Taddese, J. Hart, T.M. Antonsen, E. Ott, S.M. Anlage, *Appl. Phys. Lett.* 95, 114103 (2009).
[10] A. Peres, *Phys. Rev. A* 30, 1610 (1984).
[11] T. Gorin, T. Prosen, T.H. Seligman, M. Žnidaric, *Phys. Rep.* 435, 33 (2006).
[12] C.P. Slichter, *Principles of Magnetic Resonance*, 3rd ed., Springer-Verlag, New York 1990, p. 46.
[13] M. Fink, *Contemp. Phys.* 37, 95 (1996).
[14] M. Fink, D. Cassereau, A. Derode, C. Prada, P. Roux, M. Tanter, J.-L. Thomas, F. Wu, *Rep. Prog. Phys.* 63, 1933 (2000).
[15] G. Lerosey, J. de Rosny, A. Tourin, A. Derode, G. Montaldo, M. Fink, *Phys. Rev. Lett.* 92, 193904 (2004).
[16] G. Lerosey, J. de Rosny, A. Tourin, A. Derode, M. Fink, *Appl. Phys. Lett.* 88, 154101 (2006).
[17] S.M. Anlage, J. Rodgers, S. Hemmady, J. Hart, T.M. Antonsen, E. Ott, *Acta Phys. Pol. A* 112, 569 (2007).
[18] C. Draeger, M. Fink, *Phys. Rev. Lett.* 79, 407 (1997).
[19] B.T. Taddese, M.D. Johnson, J.A. Hart, T.M. Antonsen, E. Ott, S.M. Anlage, *Acta Phys. Pol. A* 116, 729 (2009).
[20] T.J. Ulrich, P.A. Johnson, R.A. Guyer, *Phys. Rev. Lett.* 98, 104301 (2007).
[21] J. V. Candy, D. H. Chambers, C. L. Robbins, B. L. Guidry, A. J. Poggio, F. Dowla, C. A. Hertzog, *J. Acoust. Soc. Am.* 120, 838 (2006).
[22] H. M. Pastawski, E. P. Danieli, H. L. Calvo and L. E. F. Foa Torres, *Euro. Phys. Lett.*, 77 40001 (2007).
[23] T. Gorin, T.H. Seligman, R.L. Weaver, *Phys Rev E*, 73, 015202(R) (2006).
[24] R. Schäfer, H.-J. Stöckmann, *Phys. Rev. Lett.* 95, 184102 (2005).
[25] B.Köber, U. Kuhl, H.-J. Stöckmann, T. Gorin, D.V. Savin, T.H. Seligman, *Phys. Rev. E* 82, 036207 (2010).
[26] O. Lobkis, R. Weaver, *Phys. Rev. E* 78, 066212 (2008).
[27] R. Schäfer, T. Gorin, T.H. Seligman, H.-J. Stöckmann, *New J. Phys.* 7, 152 (2005).
[28] A. H. Quazi, U.S. Patent No. 5,841,735. (Nov. 24, 1998).
[29] C. Draeger, M. Fink, *J. Acoust. Soc. Am.* 105, 611 (1999).
[30] H.L. Anderson (Editor), *Physics Vade Mecum*, American Institute of Physics, New York 1981, p65.
[31] In the preliminary publication [9], we used an empirical polynomial that accompanied the exponential amplification. The role of the polynomial is now played by a time windowing procedure described in Section IV.3.
[32] A. Papana, D. Kugiumtzis, *Int. J. Bifurcation Chaos Appl. Sci. Eng.* 19, 4197 (2009).